\documentclass{article}
\usepackage{spconf,amsmath,graphicx,subfigure, textcomp, xcolor}

\title{Detecting intracranial aneurysm rupture from 3D surfaces using a novel GraphNet approach}
\name{Z. Ma$^{\star}$ \quad L. Song$^{\star}$ \quad X. Feng$^{\star\star}$ \quad G. Yang$^{\star}$ \quad W.Zhu$^{\dagger\dagger}$ \quad J. Liu$^{\dagger\dagger}$ \quad Y. Zhang$^{\dagger\dagger}$ \quad X. Yang$^{\dagger\dagger}$ \sthanks{Correspondence to: Xinjian Yang $<$yangxinjian@voiceoftiantan.org$>$.} \quad Y. Yin$^{\star}$\sthanks{Correspondence to: Yin Yin $<$yinyin@unionstrongtech.com$>$.}}
\address{$^{\star}$ Union Strong Technology Beijing Co., Beijing, China \\ 
$^{\star\star}$ The University of Virginia, VA, USA \\
$^{\dagger\dagger}$ Beijing Tiantan hospital, Beijing, China}
\begin{document}
%

\maketitle
\begin{abstract}
Intracranial aneurysm (IA) is a life-threatening blood spot in human's brain if it ruptures and causes cerebral hemorrhage.  It is challenging to detect whether an IA has ruptured from medical images.  In this paper, we propose a novel graph based neural network named GraphNet to detect IA rupture from 3D surface data. GraphNet is based on graph convolution network (GCN) and is designed for graph-level classification and node-level segmentation.  The network uses GCN blocks to extract surface local features and pools to global features. 1250 patient data including 385 ruptured and 865 unruptured IAs were collected from clinic for experiments.  The performance on randomly selected 234 test patient data was reported. The experiment with the proposed GraphNet achieved accuracy of 0.82, area-under-curve (AUC) of receiver operating characteristic (ROC) curve 0.82 in the classification task, significantly outperforming the baseline approach without using graph based networks.  The segmentation output of the model achieved mean graph-node-based dice coefficient (DSC) score 0.88.


\end{abstract}
\section{Introduction}
\label{sec:intro} 
An intracranial aneurysm (IA) is a spot that fills of blood in arterial vessel wall.  It could cause high death or disability rate if an IA ruptures and causes subarachnoid hemorrhage (SAH).  In clinic, IAs are detected with the aid of brain vessel imaging techniques such as computed tomography angiography (CTA), magnetic resonance angiography (MRA) and digital subtraction angiography (DSA).  Immediate treatment is needed for ruptured IA once found.  However, it is challenging to detect whether an IA has ruptured from medical images.  Therefore, classifying IA's rupture is helpful for neurologists in clinic.  On the other hand, it is also helpful to obtain the shape information of IAs to analyze its physical stability from IA segmentation. 

Although medical images can provide 3-dimensional (3D) IA shape information, it is still difficult to judge whether an IA has ruptured or not visually, e.g. as seen in Figure\ \ref{fig:show_figs}.  Recently, many works focus on analyzing IA's rupture through machine learning methods using models such as logistic regression or neural networks.  The IA shape features are usually hand-engineered or automatically computed from medical images.  In this paper, we propose to extract IA shape information from their 3D surfaces.  The surface data can be easily generated from different medical image modalities using image thresholding and the Marching Cubes algorithm.  Therefore, a model built with surface data can be easily adapted to applications with different IA images such as CTA, MRA and DSA.

Deep neural network (DNN) based on image data is popular today in many fields, such as computer vision, auto driving, and face recognition, but applications using DNN on surfaces in medical domain is rare.  Here, we propose a graph based neural network called GraphNet to detect IA rupture and segment it based on 3D surface data.  GraphNet is based on a general graph convolution network (GCN) with node and edge information as the input and node-level segmentation or graph-level classification as the outputs.  Our GraphNet includes multiply GCN blocks that learn the local representations from 3D surfaces, and a global pooling layer that pools local representations to global features.   The clinical and hand-engineered morphological features can be aggregated to the global features to further increase the model performance.  To the best of our knowledge, this is the first graph based neural network designed to analyze surface data in medical domain.

\begin{figure*}[htbp]
\centering
\includegraphics[width=16cm]{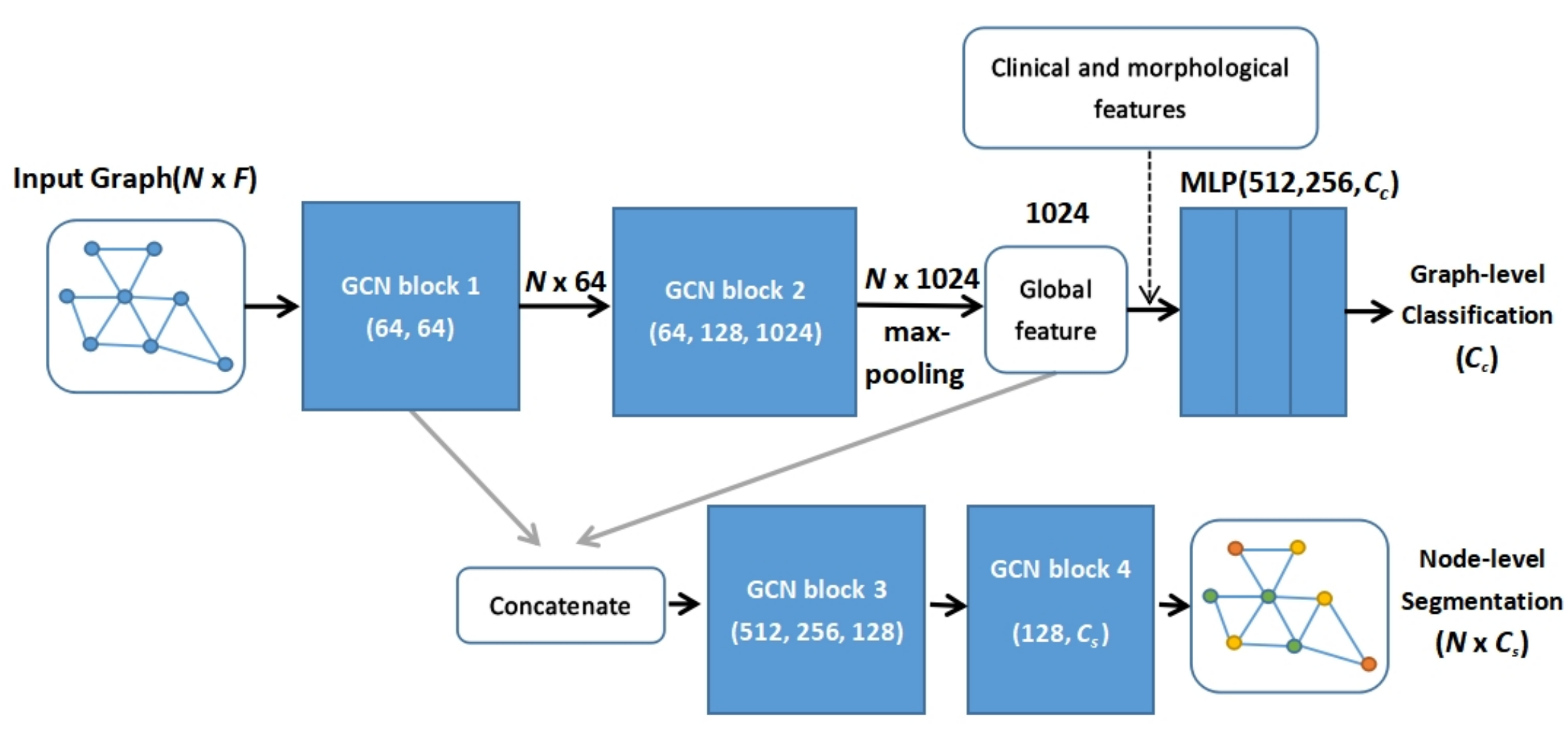}
\caption{
    GraphNet structure. The classification network takes input graph with $N$ nodes, $F$ features and an $A$ matrix.  Then it applies two GCN blocks and aggregate local features to global features by a max-pooling layer.  Each GCN block contains multiple GCNs.  The outputs are graph-level classification scores for $C_c$ classes.  The segmentation network takes the same input graph, and concatenates the global features to its local features and outputs node-level segmentation for $C_s$ classes.}
\label{fig:GraphNet}
\end{figure*}

\section{METHODS}
\label{sec:majhead}

A graph $G(n,v)$ is composed of $N$ nodes $n$ and a set of edges $v$ that can be represented as a $N \times N$ adjacency matrix $A$.  An edge is a connection between two neighboring nodes, and a node may contain $F$ features.  Graph can be irregular.  A graph may have variable number of unordered nodes, and each node in that graph has different number of neighborhoods connected by edges.   Images, point clouds and surfaces can be looked as special cases of graphs, where images have ordered nodes (pixels) and each node has the same number of neighboring nodes; point clouds only have unordered nodes without edges; and 3D triangulated surfaces have nodes from vertices and edges from triangle edges.  For surface data, the features associated to each node can be the node's Cartesian coordinates.  


\subsection{GCN block}
\label{ssec:subhead}

Graph convolutional network(GCN)\ \cite{Kipf2016Semi} is a popular approach to analysis graphs.  A GCN with an input graph as a $N \times F$ feature matrix $X$ and an adjacency matrix $A$ can be represented as:
\begin{equation}
    f(A, X) = \sigma(\Tilde{D}^{-1/2}\Tilde{A}\Tilde{D}^{-1/2}XW)
\end{equation}
and
\begin{equation}
   \Tilde{A}=A+I
\end{equation}

where $W$ is a $F\times K$ weight matrix with hidden size $K$ that to be learned. $\sigma$ is an activation function e.g. ReLU.  $I$ is an identity matrix. $\Tilde{D}$ is the diagonal degree matrix of $\Tilde{A}$, which is used to normalize rows of $\Tilde{A}$. 

A GCN block contains a series of $L$ connected GCNs:
\begin{equation} \label{eq:GCN_block}
    H^{l+1} = f(A, H^{l}) \qquad 0\leq{l}\leq{L}
\end{equation}

For example, a simple node-level classification model with only one GCN block has $H^0=X$ and $H^L$ is an output matrix with dimension $N \times K^L$ where $K^L=C$ is the number of output classes per node.

\subsection{GraphNet design}
\label{ssec:subhead}

The proposed GraphNet contains multiple GCN blocks can be seen in Figure\ \ref{fig:GraphNet}.  GraphNet can be divided into two parts, the classification network and the segmentation network.  These networks share part of the structures.  The classification network outputs graph-level classification scores.  The segmentation network outputs classification scores for each node for node-level segmentation.  On the classification side of the network, the input $N$ nodes are encoded to $N \times 1024$ features through two GCN blocks with hidden sizes $(64, 64)$ and $(64, 128, 1024)$ respectively.  Those features are further pooled to global features by a max-pooling layer to $1 \times 1024$.  The $1 \times 1024$ features input to a multi-layer perceptron (MLP) with hidden sizes $(512, 256, C_c)$, and outputs a per-class score for graph-level classification.  On the segmentation side of the network, the local and global features are concatenated to a $N \times 1088$ feature matrix, so that each of the node contains both local and global information for segmentation.  The $N \times 1088$ matrix is decoded to node-level scores through another two GCN blocks with hidden sizes $(512, 256, 128)$ and $(128, C_s)$ correspondingly. 

The GraphNet design is similar to PointNet\ \cite{Charles2017PointNet}, a powerful network to process point clouds.  However, the distinct difference between our GraphNet and PointNet is that we designed GCN blocks to include node relation information in GraphNet.  As such, PointNet can be looked as a special case of GraphNet without node connection information.  In another word, PointNet is equivalent to GraphNet with $A=0$.

\subsection{Clinical and morphological features aggregation}
\label{ssec:subhead}

Embedding different sources of medical knowledge and information into a model is very useful for machine learning based medical applications.  Aggregating additional medical features such as clinical features from medical report or hand-engineered shape morphological features into GraphNet can be achieved by concatenating those features directly to the $1024$ global feature vector.

\section{EXPERIMENTS}
\label{sec:print}

\subsection{Dataset}
We collected 3D brain vessel surfaces of 1250 patients from clinic following related regulations.  The surfaces were generated from patients' brain DSA sequences using the thresholding and Marching Cubes algorithms.  Each patient has only one generated surface and contains either one unruptured IA (UIA) or ruptured IA (RIA). The UIA and RIA ground truth was determined by multiple clinical experts by observing the corresponding DSA sequences with the aid of additional CT scans providing hemorrhage information.  The IA surfaces were manually segmented using a 3D annotation tool for all the patients.  The 1250 IA surfaces include 865 UIAs and 385 RIAs.  For each patient, 10 clinical information were extracted from the medical reports and quantized as clinical features including patient's age, sex, and the histories of drinking, smoking, hypertension, other symptoms, coronary heart disease, diabetes, family diseases and hyperlipidemia.  In addition, 25 IA shape morphological measurements were computed based on the prior knowledge from clinical publications.  These features measure the IA's location and basic shape information such as IA located vessel, and IA's diameter-to-width-ratio, diameter, vessel-mean-diameter, volume, width, etc.  The dataset we collected has IA mean diameter $4.87mm$ with the largest IA diameter $27.8mm$ and the smallest diameter $1.31mm$.  The mean patient age is 55.3 with the oldest age 86 and the youngest age 13.

Only the manually segmented IA surfaces and small portions of surrounding vessel surfaces were used for experiments to mimic the situation that the IA regions in question had been pre-selected by neurologists in use.  We limited each IA and surrounding vessel surface region to contain 1024 vertices or graph nodes.  The average IA and surrounding vessel nodes are roughly equal.  Each node's 3D Cartesian coordinates $(x, y, z)$ were used as features.   The triangulation connection information of each surface was converted into a $1024\times1024$ adjacency matrix.  1016 patients were randomly selected to train the proposed GraphNet and the remaining 234 patients were used to evaluate the performance of the model. 

\subsection{Experiments design}

During training, the 1016 $1024\times3$ graph feature matrices and their corresponding $1024\times1024$ adjacency matrices were input to the GraphNet.  The graph-level labels for UIA or RIA and node-level manual segmentations were used as targets.  The batch size was set to 32, and the initial learning rate was set to 0.001 with learning rate exponential decay rate of 0.7.  Cross entropy loss function was used for the classification part, and graph-node-based dice coefficient (DSC) loss function was for the segmentation part.   Those two losses were added together with equal weights, and optimized with an Adam optimizer.  To avoid over-smoothing of features due to multiplying $A$ matrices multiple times, we set $A=0$ for $l>0$ in Equation\ \ref{eq:GCN_block} for each GCN block. The whole network was trained for 100 epochs, and the performance was evaluated from the test graphs.

We measured the mean classification accuracy for the test graphs, and plotted the receiver operating characteristic (ROC) curves for different experiment settings below. Furthermore, we measured the mean DSC for graph nodes and plotted examples of the segmented test surfaces. 

The experiments we measured were:
\begin{itemize}

\item \emph{Baseline:} The 10 clinical and 25 hand-engineered morphological features were directly input to a 3-layer MLP classifier for UIA and RIA classification.  The hidden sizes for MLP layers are identical to the last MLP we used in GraphNet for graph-level classification.  We also tried other classifiers like logistic regression and Random Forest, but the performance was similar or worse than MLP.  

\item \emph{Baseline+PointNet:} A PointNet was trained to extract shape information using only vertices of the 3D surfaces.  The PointNet we used has similar structure to the proposed GraphNet but only contains MLP layers instead of GCN blocks and does not have adjacency matrix input.  We trained and optimized the PointNet the same way as GraphNet with clinical and morphological features concatenated to the global features.

\item \emph{Baseline+GraphNet:} The proposed GraphNet was trained with clinical and morphological features as described above.

\end{itemize}

\section{RESULTS}
\label{sec:page}

\begin{figure*}[htpb]
    \centering
    \subfigure[]{
    	\label{fig:r1_real}
    	\includegraphics[height=0.175\textwidth]{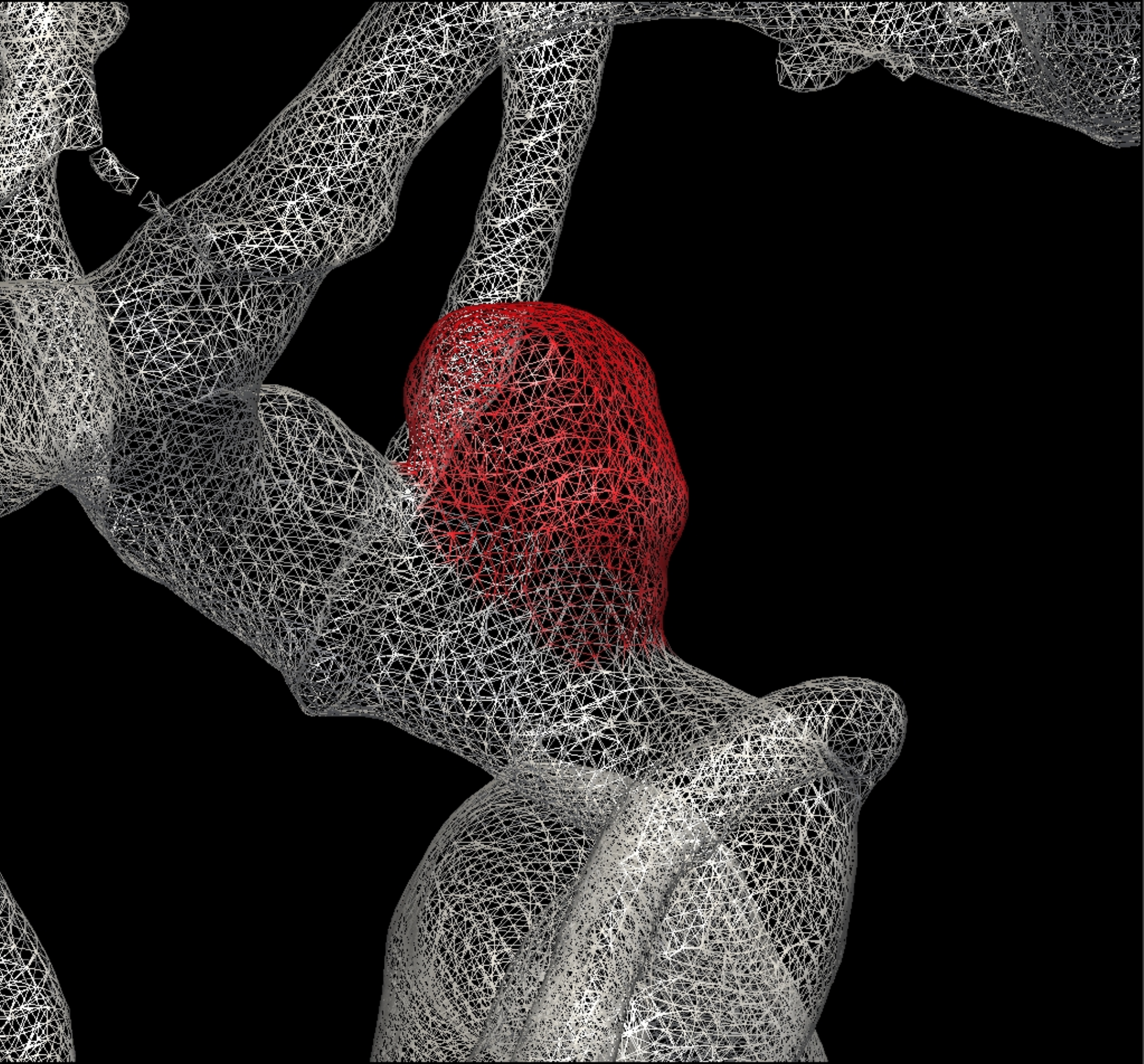}}
    \subfigure[]{
    	\label{fig:r2_real}
    	\includegraphics[height=0.175\textwidth]{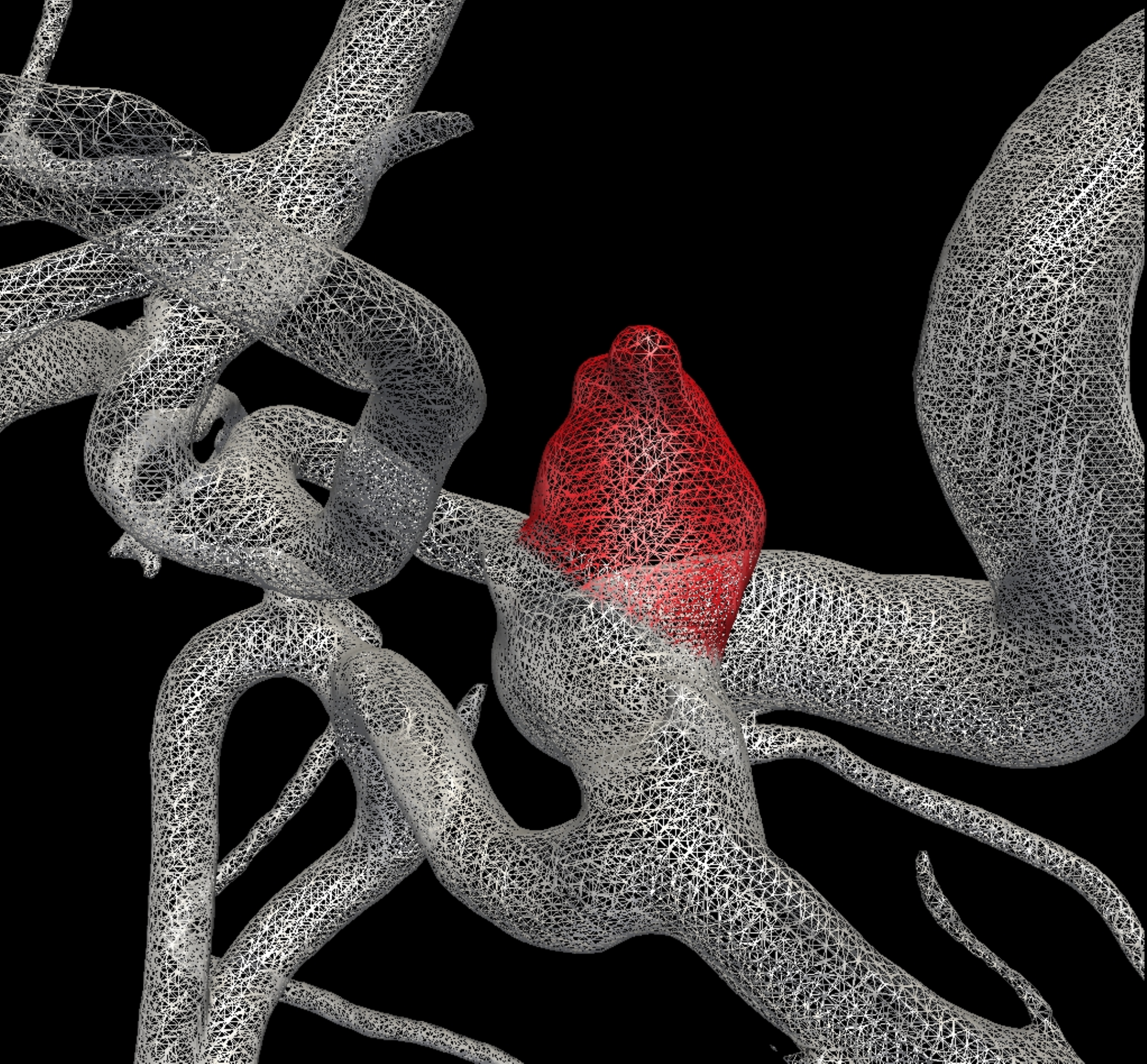}}
    \subfigure[]{
    	\label{fig:u1_real}
    	\includegraphics[height=0.175\textwidth]{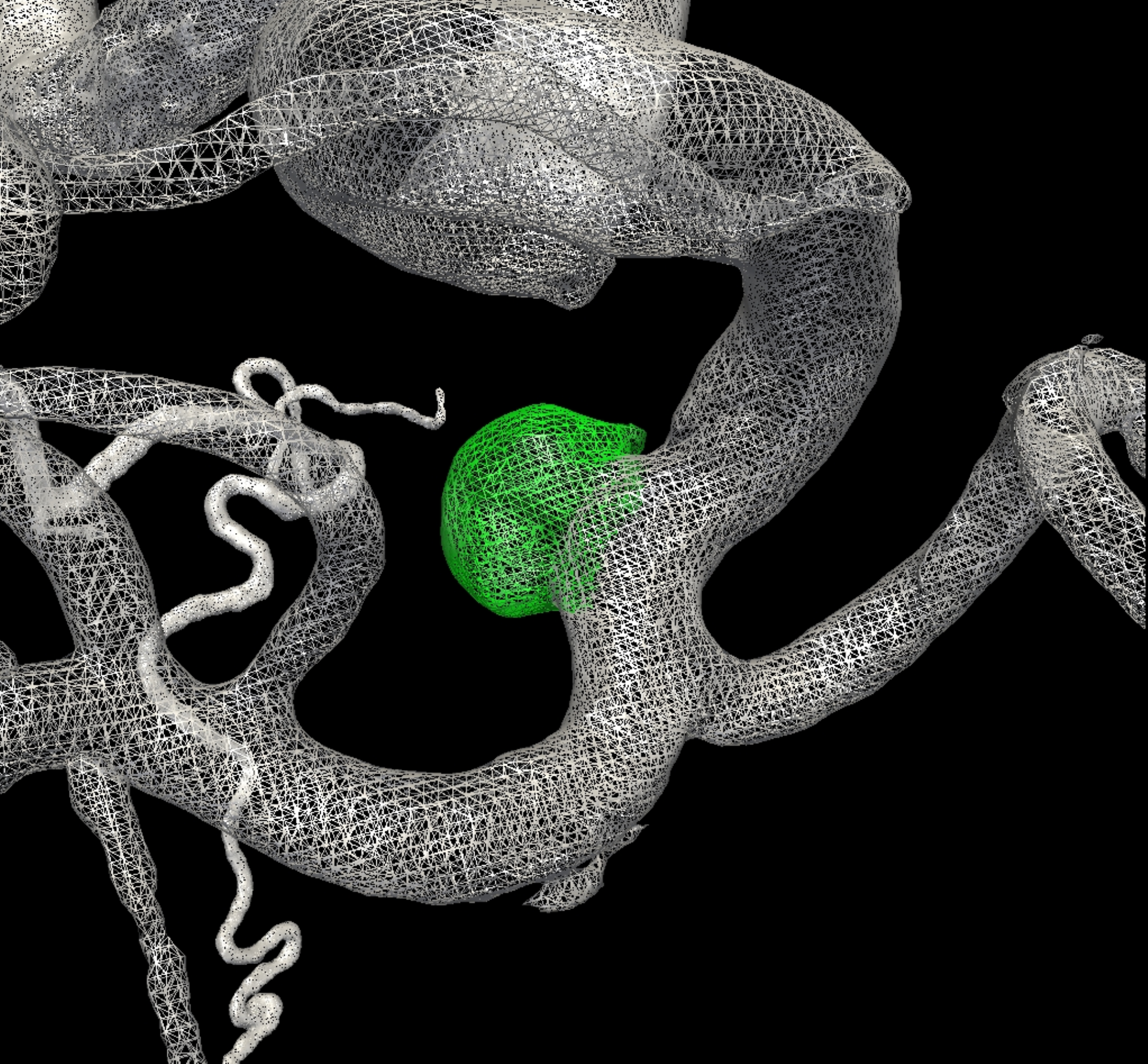}}
    \subfigure[]{
    	\label{fig:u2_real}
    	\includegraphics[height=0.175\textwidth]{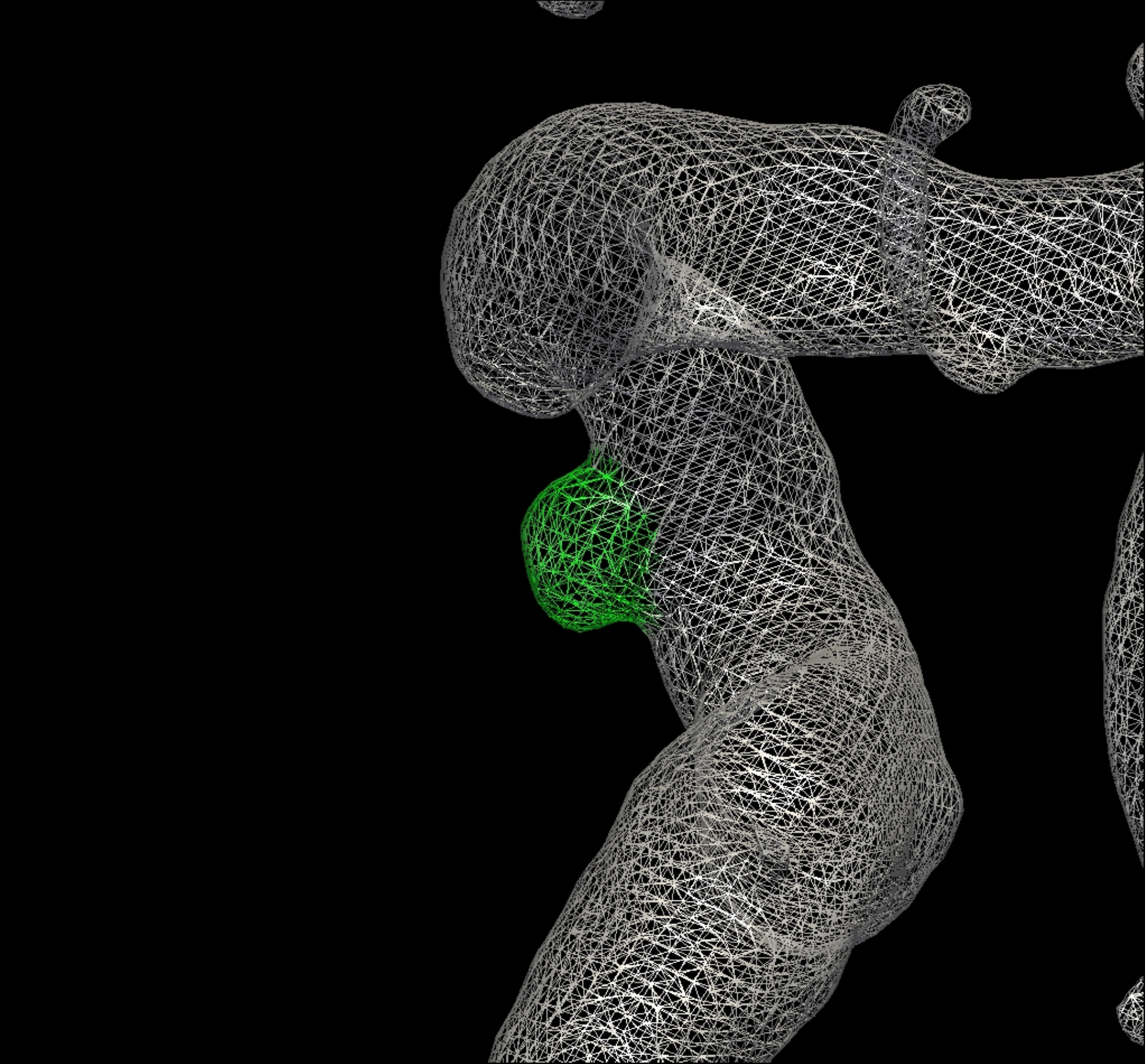}}
    \subfigure[]{
    	\label{fig:u2_real}
    	\includegraphics[height=0.175\textwidth]{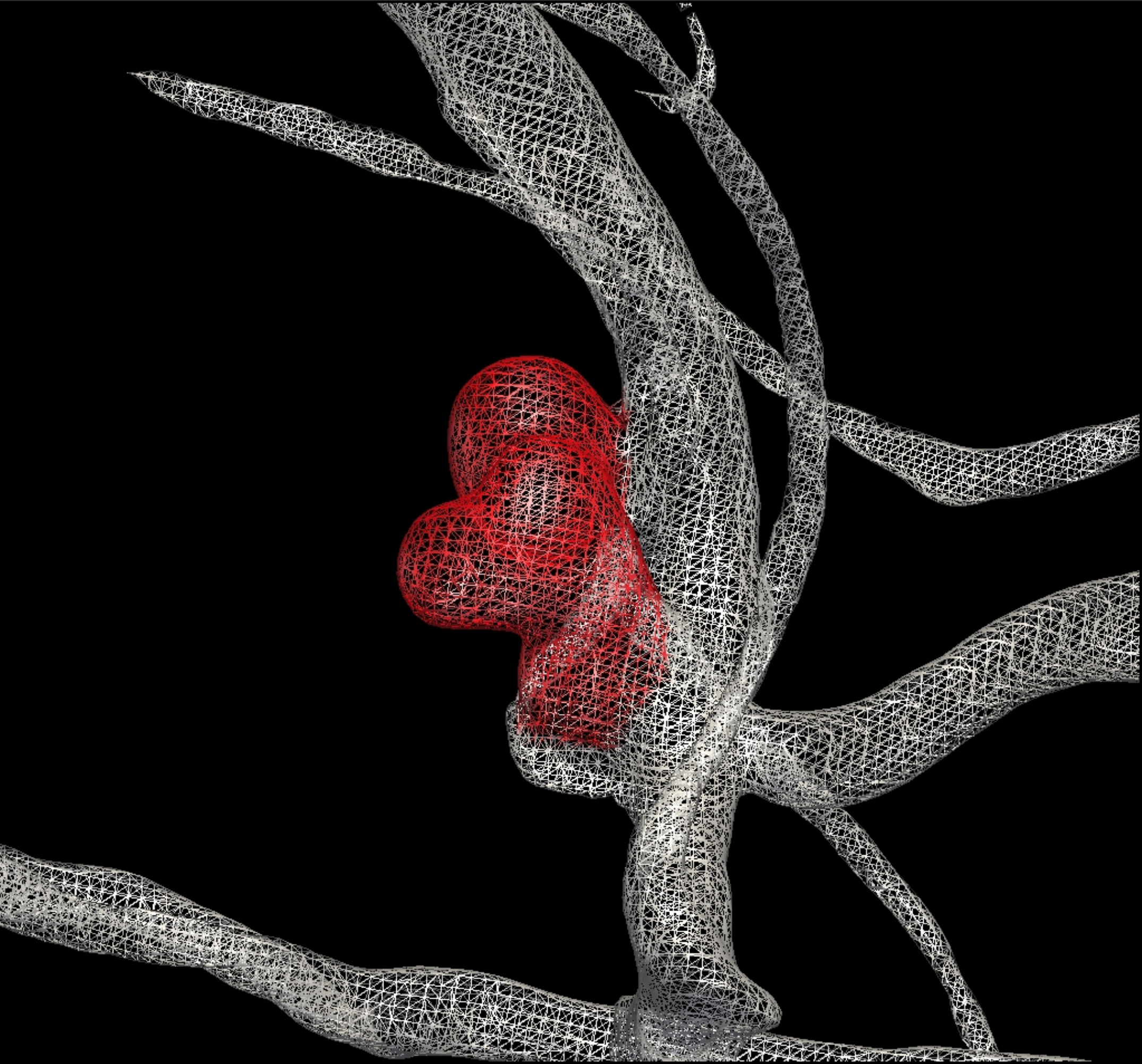}}
    \subfigure[]{
    	\label{fig:r1_pred}
    	\includegraphics[height=0.175\textwidth]{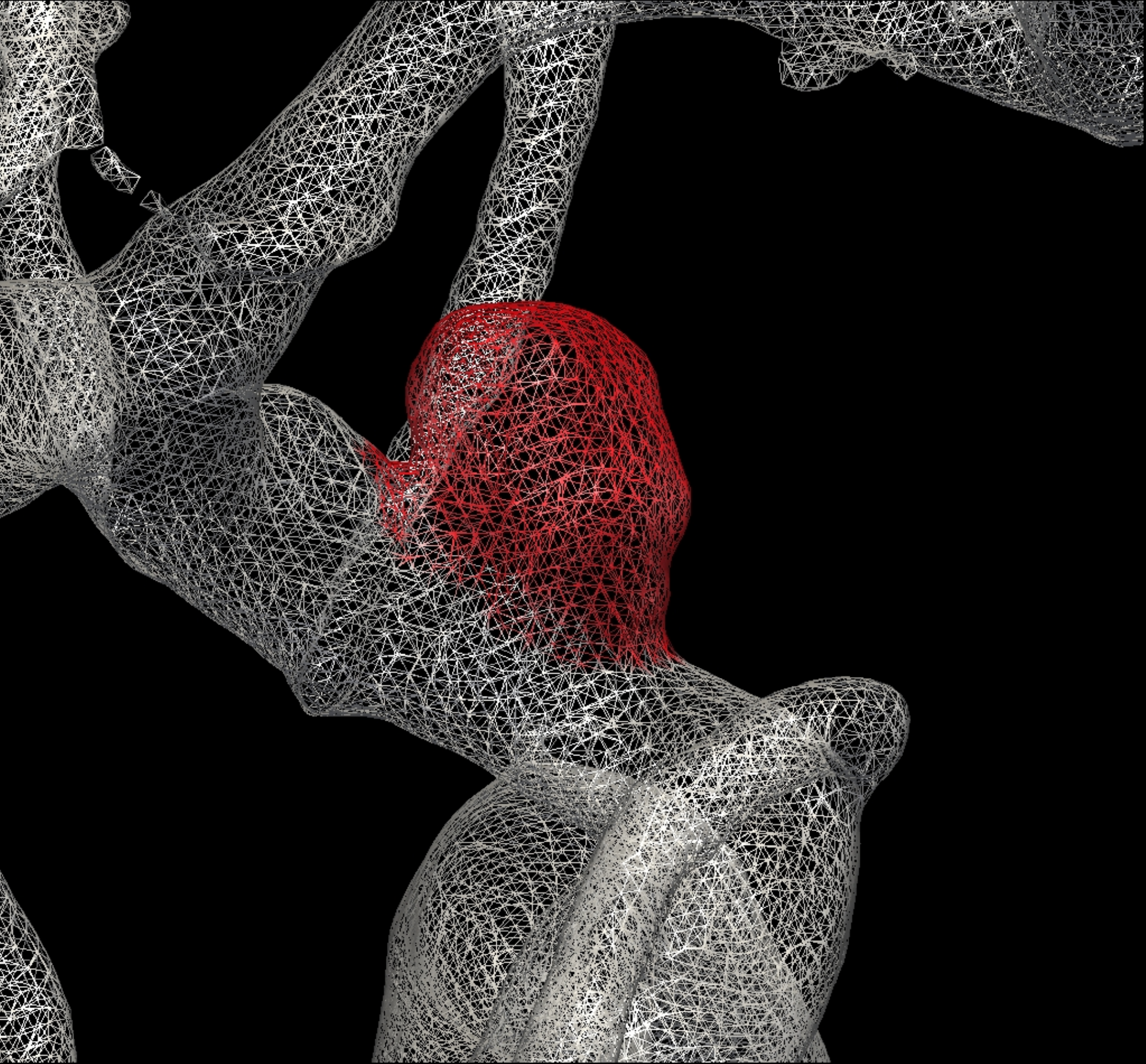}}
    \subfigure[]{
    	\label{fig:r2_pred}
    	\includegraphics[height=0.175\textwidth]{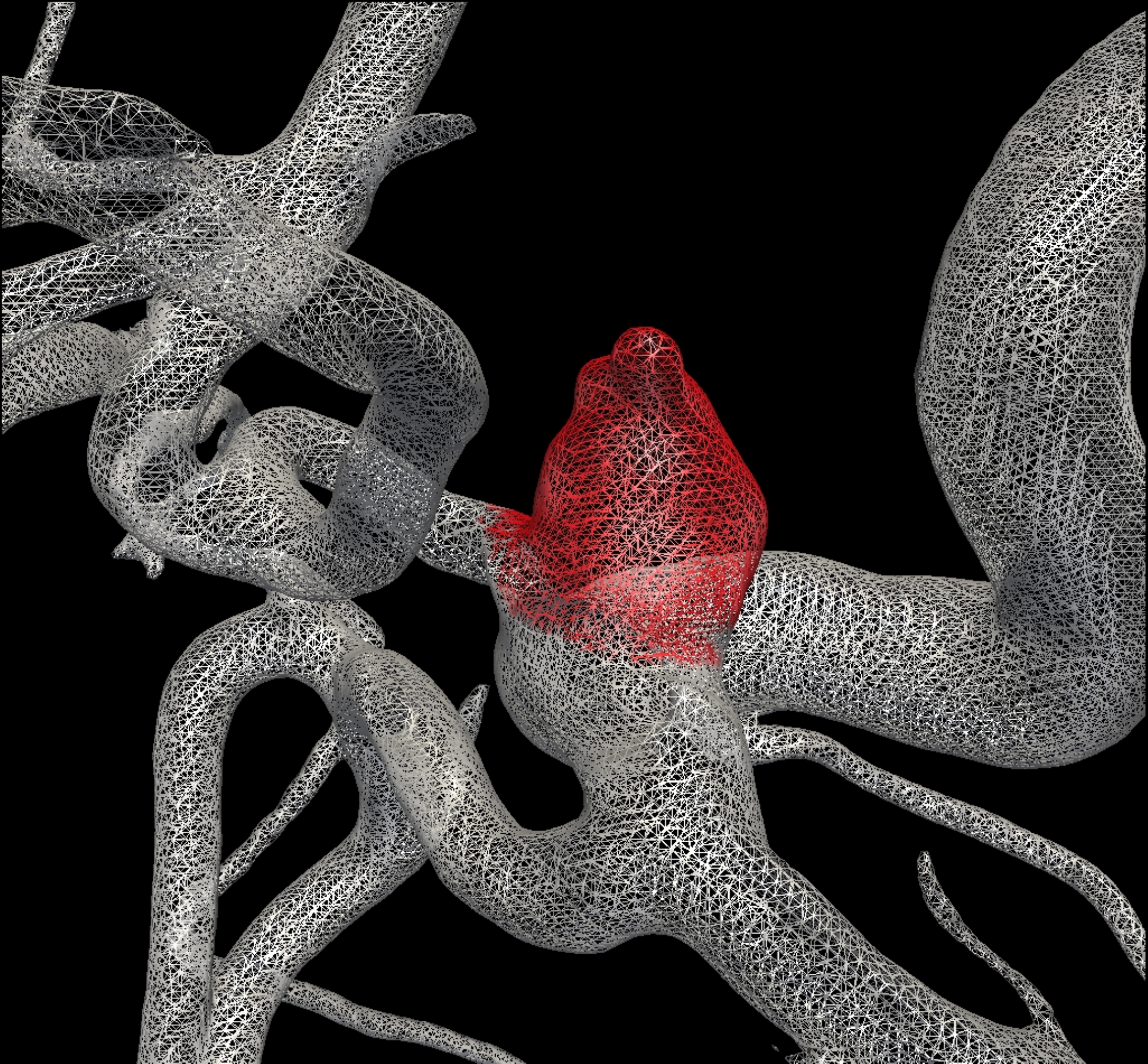}}
    \subfigure[]{
    	\label{fig:u1_pred}
    	\includegraphics[height=0.175\textwidth]{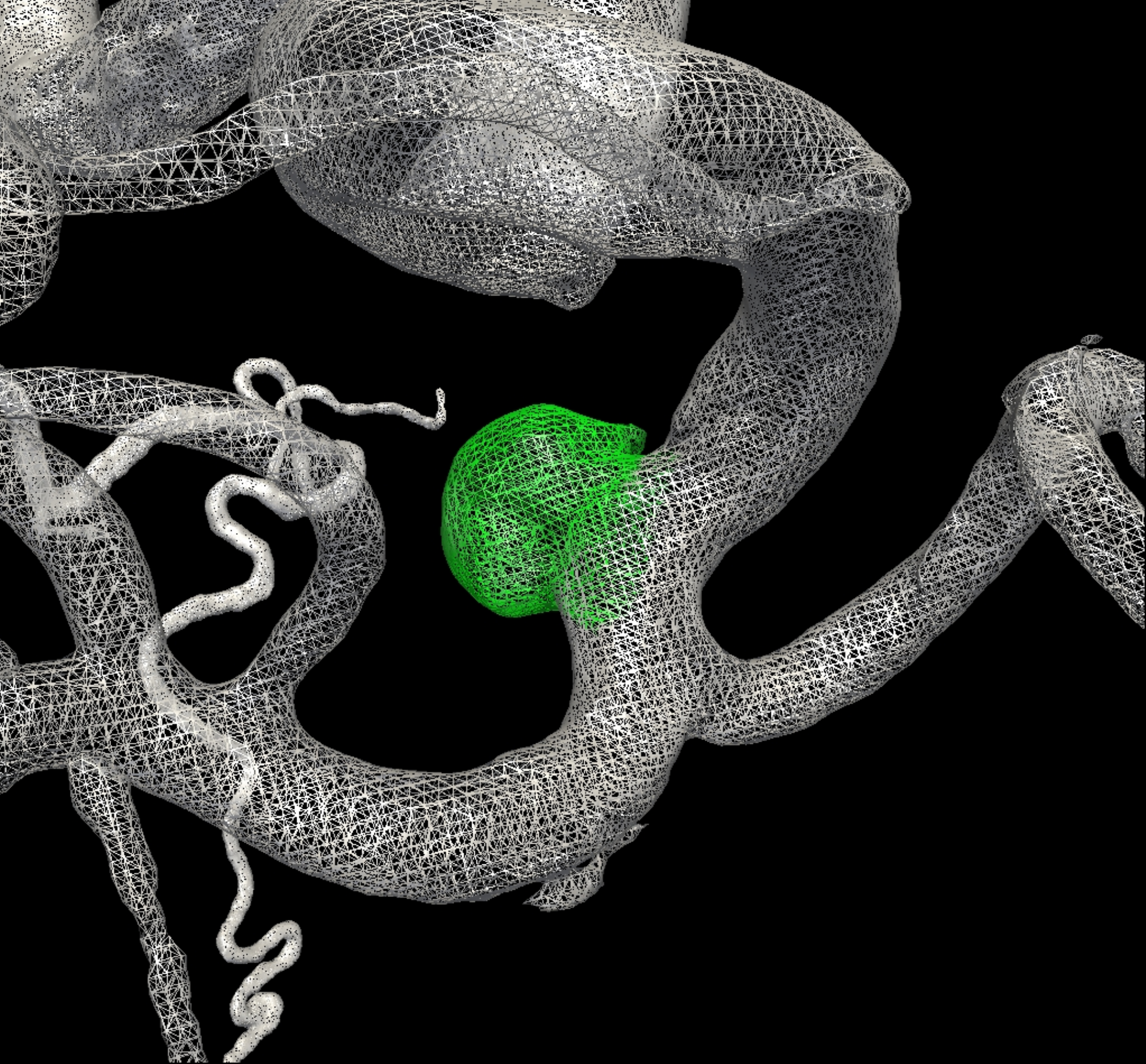}}
    \subfigure[]{
    	\label{fig:u2_pred}
    	\includegraphics[height=0.175\textwidth]{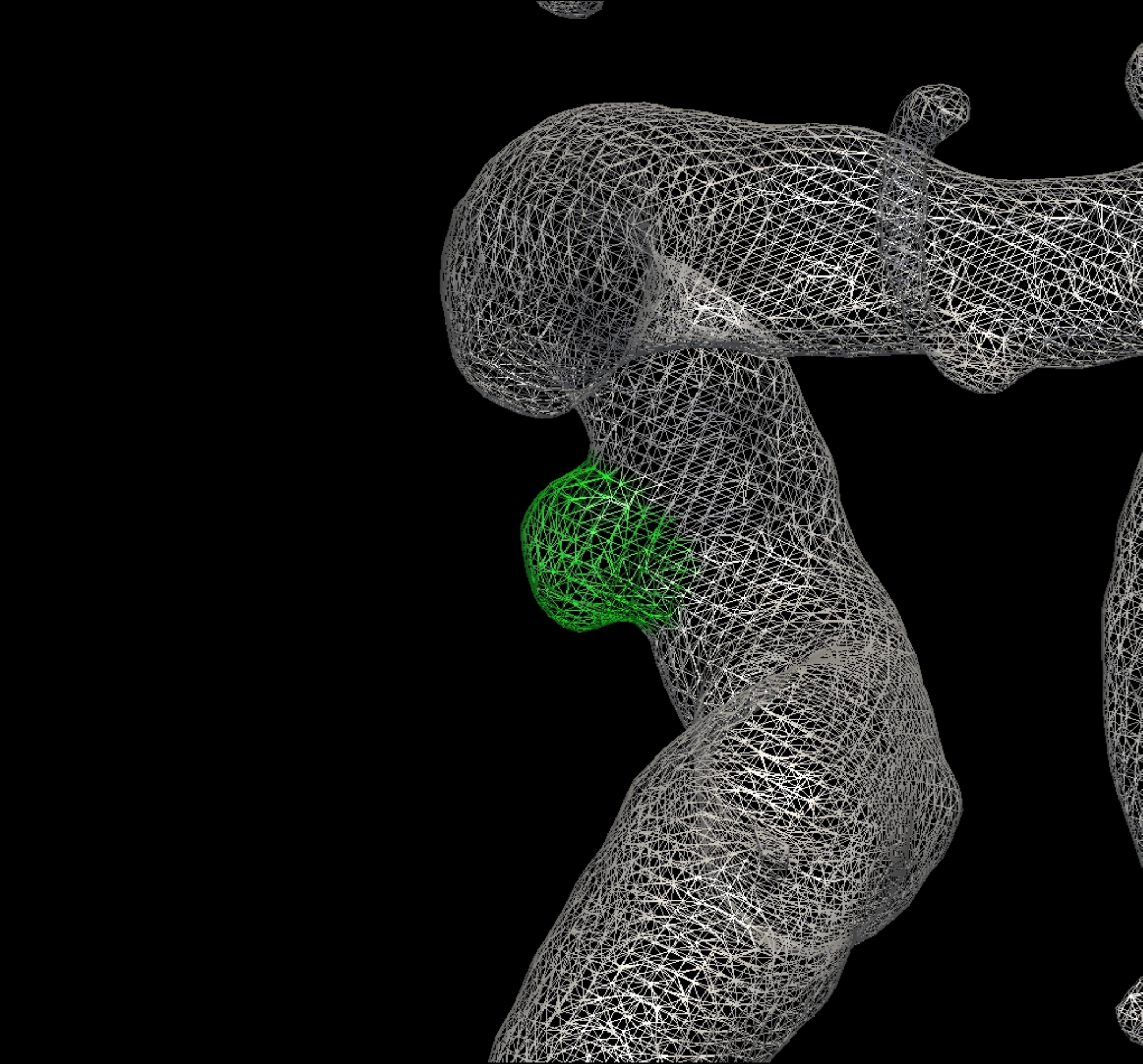}}
    \subfigure[]{
    	\label{fig:u2_pred}
    	\includegraphics[height=0.175\textwidth]{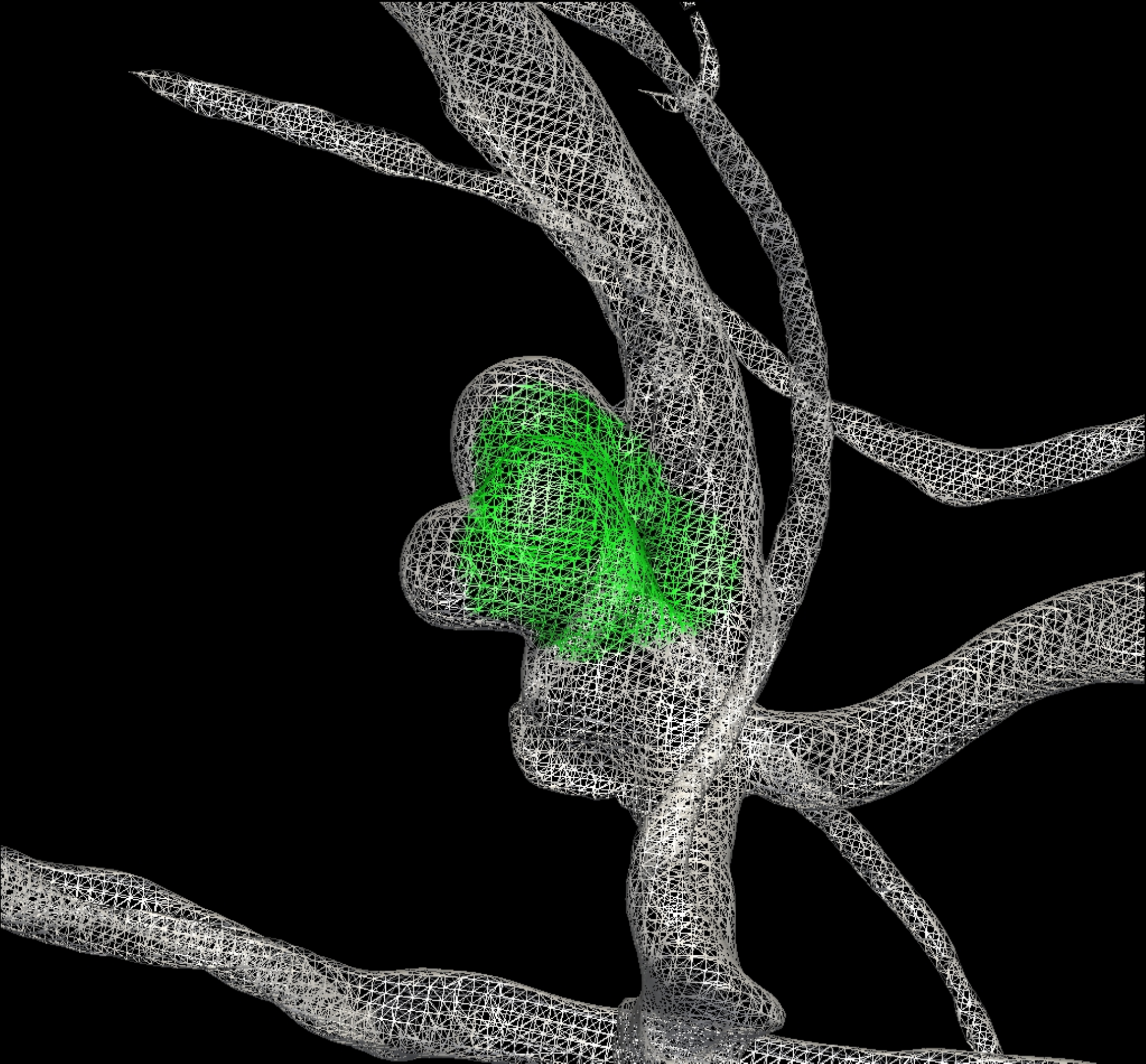}}
    \caption{
    Examples of the model outputs. Red color represents RIAs and green color is UIAs.  The upper row is the ground truth labels and the lower row shows classified and segmented results correspondingly. The first four columns are correctly classified results and the last column shows a failed case. }
     \label{fig:show_figs}
\end{figure*}

\begin{figure}[htpb]
    \centering
    \includegraphics[width=7.cm]{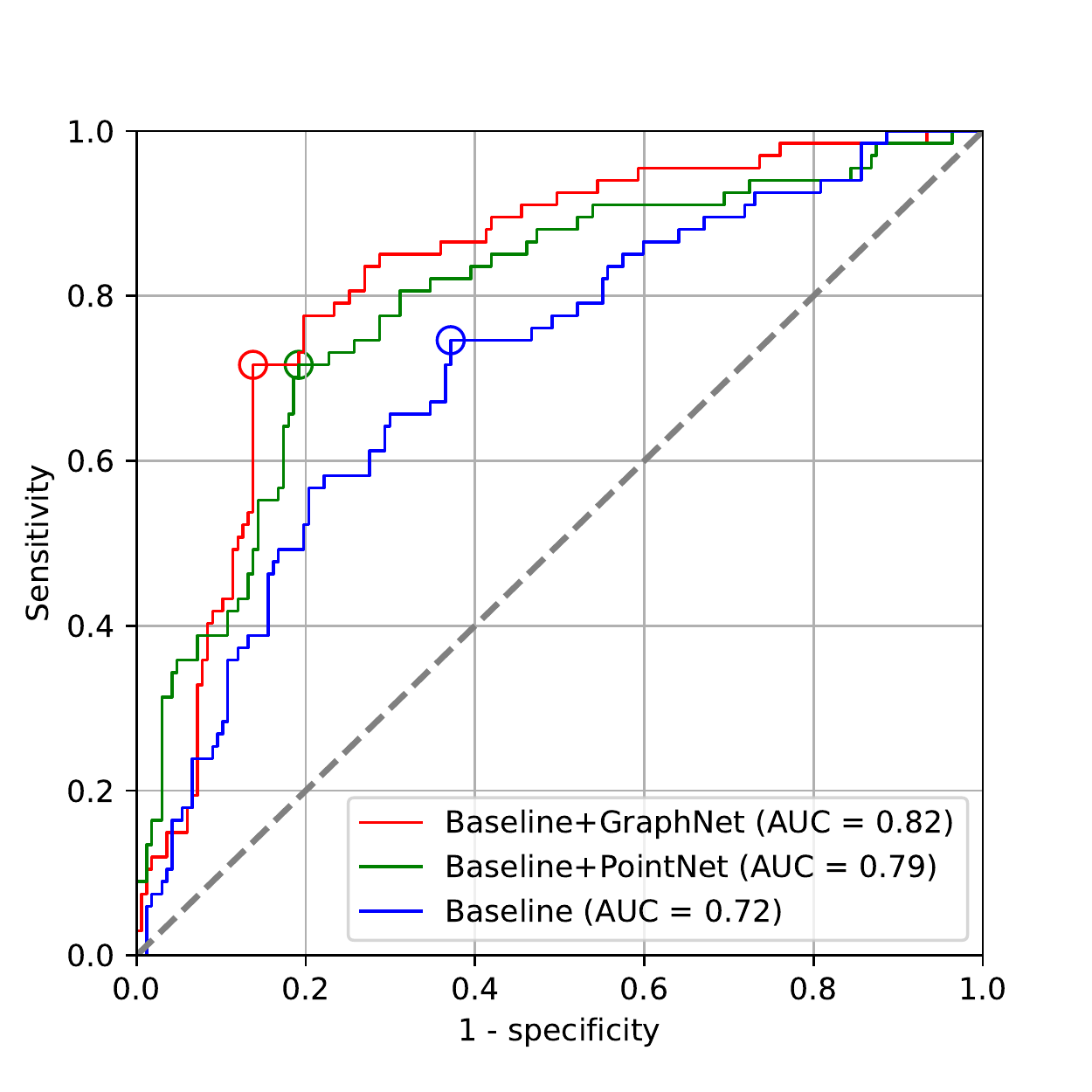}
    \caption{
    ROCs of the experiments Baseline, Baseline+PointNet and the proposed Baseline+GraphNet with the corresponding Youden's index determined operating points as circles.}
    \label{fig:result}
\end{figure}

Figure\ \ref{fig:result} shows the ROC curves over the 234 test patient data.  The area-under-curves (AUCs) for experiments Baseline, Baseline+PointNet and Baseline+GraphNet are 0.72, 0.79 and 0.82 respectively.  The accuracies at the Youden's index determined operating points on the three experiment ROCs are 66.2\%, 78.2\%, and 82.1\%.   The operating points are determined by maximizing $sensitivity + specificity - 1$ for each ROC, and are placed at sensitivities 0.746, 0.717 and 0.717, specificities 0.629, 0.808 and 0.862 respectively for Baseline, Baseline+PointNet and Baseline+GraphNet, labeled as circles in Figure\ \ref{fig:result}.  The mean graph-node-based DSC for the Baseline+GraphNet segmentation network is 0.88.  Four correctly classified and one failed test surfaces are provided in Figure\ \ref{fig:show_figs} with their corresponding segmentations. The average prediction time for each IA is $7.5ms$ on a contemporary workstation with Titan V GPUs, which shows the potential of real-time usage in clinic situations. 

\section{CONCLUSIONS}

In this paper, we proposed a novel graph based neural network called GraphNet to extract shape features directly from 3D surfaces.  
GraphNet uses local and global features for graph-level classification and node-level segmentation.  We used proposed GraphNet with clinical and hand-engineered morphological features to classify UIAs and RIAs and provided their segmentations based on a large brain vessel dataset collected from clinic.  The reported accuracy and the AUC of the ROC using the GraphNet outputforms baseline approaches without using graph based networks.  We plan to collect more patient data and further improve GraphNet with modern variations of GCN e.g. graph attention and pooling\ \cite{Lee19} in the future.





\bibliographystyle{IEEEbib}
\bibliography{GraphNetReferences}

\end{document}